\documentclass{aa}
\usepackage{graphicx}

\begin{document}

\newcommand{\etal}{{ et al. }} 
\newcommand{\sy}{Seyfert }
\newcommand{\be}{\begin{equation}}
\newcommand{\ee}{\end{equation}}
\newcommand{\msun}{{\rm M_{\odot }}}

   \title{On black hole masses, radio-loudness and bulge luminosities of Seyfert 
	  galaxies}
   \author{Xue-Bing Wu
          \inst{1,2}
          \and
          J. L. Han \inst{2}
          }

   \offprints{Xue-Bing Wu}

   \institute{\inst{1} Department of Astronomy, Peking University,  Beijing 100871, China\\
		\inst{2} National Astronomical Observatories, Chinese Academy of Sciences, 
	Beijing 100012, China\\
             email: wuxb@bac.pku.edu.cn; hjl@bao.ac.cn} 

   \date{Received 8 June 2001 / Accepted 19 September 2001}

   \abstract{
We estimated black hole masses for 9 Seyfert 1 and 13 Seyfert 2 
galaxies in Palomar and CfA bright Seyfert samples using the tight
correlation between black hole mass and bulge velocity dispersion. 
Combining other 13 Seyfert 1s and 2 Seyfert 2s  in these samples   
but with black hole masses 
measured recently by reverberation mapping and stellar/gas 
dynamics, we studied 
the correlations of black hole masses with
radio loudness and bulge luminosities for a sample of 37 \sy galaxies.
We found that if 
radio-loudness is measured using the optical and radio luminosities of 
the nuclear components, the black hole masses
of radio-loud Seyfert 1s tend to increase with the radio-loudness. 
The black hole masses of all Seyfert galaxies  increase
with the radio power, but  Seyfert galaxies have larger radio powers than
nearby galaxies with the same black hole masses. 
In addition, the correlation between black hole masses 
and bulge V-band luminosities for Seyfert galaxies is  consistent with 
that found for quasars and normal galaxies. The combined sample of 37 Seyfert galaxies,
15 quasars and 30 normal galaxies  suggests a possible universal nonlinear relation
 between black hole and bulge masses, 
${\rm M_{\rm BH} \propto M}_{\rm bulge}^{1.74 \pm 0.14}$,  %%
which is slightly steeper than that found recently
by Laor (2001) for a smaller sample. This nonlinear relation is supported 
by a larger sample including 65 \sy galaxies. The different 
${\rm M_{\rm BH}/ M}_{\rm bulge}$ ratio for galaxies with different bulge  %%
 luminosities or different
black hole masses may be explained by this relation. 
These results are consistent with some theoretical implications and are important for  
understanding the nature of radio emissions
and the formation and evolution of supermassive black holes and galaxies.
   }
\titlerunning{Black hole masses of \sy galaxies}

   \maketitle
%________________________________________________________________

\section{Introduction}

Supermassive black holes (SMBHs), with masses in the range of $10^6$ to $10^9 \rm M_\odot$, 
have been suggested to exist in the center of quasars 
and active galactic nuclei (AGNs). Accretion onto these SMBHs may account for the 
huge power
 of these energetic objects (Lynden-Bell 1969; Rees 1984). Recently, the masses of central 
objects in 20 Seyfert galaxies and 17 nearby quasars have been measured with the 
reverberation 
mapping technique (Ho 1999; Wandel, Peterson \& Malkan 1999; Kaspi \etal 2000), which confirmed
the existence of  SMBHs in the center of these objects. On the other hand, a lot of 
observations using gas and stellar dynamics indicated that SMBHs probably also exist 
in the center 
of our Galaxy (Ghez \etal 1998; Genzel \etal 1997) and in the nuclei of many normal 
galaxies 
(Kormendy \& Richstone 1995; Magorrian \etal 1998; Kormendy \& Gebhardt 2001). However, 
the apparent inactive 
feature of  normal galaxies seems to suggest that the central engines of these 
galaxies may be different from those in quasars and AGNs. It was suspected that
advection-dominated accretion flows (ADAFs; see Narayan, Mahadevan \& Quataert 1998 and
Kato, Fukue \& Mineshige 1998 
for reviews) with 
very low accretion rate and low radiative efficiency may exist in the nuclei of normal 
galaxies (Fabian \& Rees 1995; Di Matteo \& Fabian 1996), while quasars
and most AGNs probably host the standard geometrically thin accretion disks (Shakura \&
Sunyaev 1973; Novikov \& Throne 1973) with higher accretion rate.

One interesting result found recently in the searches of SMBHs in nearby galaxies is the 
correlation of black hole masses with the properties of galactic bulges. Although with 
large scatters, the black hole masses
 seem to correlate with  bulge luminosities (Kormendy
1993; Kormendy \& Richstone 1995; Magorrian \etal 1998). This also leads to the finding 
that the black hole mass, $\rm M_{\rm BH}$, is possibly proportional to the bulge mass, 
${\rm M}_{\rm bulge}$,  %%
though the mass ratio
found by different authors was different, in the range of $0.2\%$ to $0.6\%$
(Kormendy \& Richstone 1995; Magorrian \etal 1998; Ho 1999).  
Laor (1998; 2001) recently found that
some nearby quasars and Seyfert galaxies follow nearly the same $\rm M_{\rm BH}$-bulge 
luminosity relation as normal galaxies, and suggested a universal nonlinear 
relation, ${\rm M_{\rm BH} \propto M}_{\rm bulge}^{1.54\pm0.15}$, for both normal and active  %%
galaxies. This means that the ${\rm M_{BH} / M}_{\rm bulge}$ ratio is not constant for galaxies %%
with different bulge luminosities. Recently, a significantly tight correlation
of $\rm M_{\rm BH}$ with the bulge velocity dispersion $\sigma$ was also found for nearby
galaxies (Gebhardt \etal 2000a; Ferrarese \& Merritt 2000).
More recent studies
indicated that  
11 Seyfert galaxies  with $\rm M_{\rm BH}$ measured by reverberation mapping 
follow the same $\rm M_{\rm BH}$-$\sigma$ relation as for normal galaxies 
(Gebhardt \etal 2000b; Ferrarese \etal 2001),  
implying another  possible universal relation for both normal and active galaxies.
The correlations of the SMBH mass with the properties of the galactic
bulge strongly suggest a tight connection between the formation and evolution of the
SMBH and galactic bulge, though the nature of this connection is still in debate
(Haehnelt \& Rees 1993; Haiman \& Loeb 1998; Silk \& Rees 1998; Haehnelt, Natarajan \& Rees
1998; Kauffmann \& Haehnelt 2000; Adams, Graff \& Richstone 2001)

Nonthermal radio emissions of quasars and AGNs  are believed to 
be probably produced by relativistic electrons that are powered by jets (Begelman, Blandford \& 
Rees 1984; Blundel \& Beasley 1998). Similar radio emissions have been  detected in 
the nuclei of normal elliptical galaxies (Sadler, Jenkins \& Kotanyi 1989) and   also in 
some spiral galaxies (Sadler \etal 1995). Recently, some studies have indicated that the 
radio power may be directly correlated with the black hole mass. Franceschini, 
Vercellone \& Fabian (1998) found a very tight relation between black hole mass and 
radio power in a small sample of nearby mostly non-active galaxies. McLure \etal (1999)
estimated the black hole masses for a sample of AGNs using $\rm M_{\rm BH}$-bulge 
mass relation found by Magorrian \etal (1998) and noted they follow the same
correlation with radio power as found by Franceschini \etal (1998). However,
Laor (2000) recently argued that the $\rm M_{\rm BH}$ and radio power of 
a sample of 87 $z<0.5$ Palomar-Green (PG) quasars (Schmidt \& Green 1983;
Boroson \& Green 1992) do not follow the tight correlation suggested by Franceschini
\etal (1998), because quasars usually have over 100 times larger radio power than normal galaxies
at a given $\rm M_{\rm BH}$. He suggested that the larger scatters of radio power at
a given $\rm M_{\rm BH}$ may be simply due to the different levels 
of overall continuum luminosity of different objects. Laor (2000) 
 also noted that the radio-loud quasars seem to host more massive black 
holes than radio-quiet quasars. Very recently, Ho \& Peng (2001) studied the 
radio-loudness of bright Seyfert 1 galaxies using the nuclear
 radio and optical luminosities, and suggested 
that the majority of Seyfert 1 nuclei in their sample are essentially radio loud. 
Therefore, it is useful and feasible to check 
 if these radio-loud Seyfert  nuclei  host more massive black holes than  
radio-quiet ones 
 and if the Seyfert galaxies still follow 
the same correlation between the radio power and black hole masses of 
 nearby galaxies and quasars (Franceschini \etal 1998; Laor 2000).
In this paper we will try to derive the SMBH masses for a number of Seyfert galaxies and
study their correlations with radio power and bulge luminosities. 
Further studies of these 
correlations will probably provide some important clues to help us understand the 
nature of radio emissions in active and normal galaxies and the formation 
and evolution of SMBHs and galaxies.

\section{Estimating the SMBH masses of \sy galaxies} %%

Currently the SMBH masses
of a few weak AGNs have been well measured by stellar dynamics, ionized gas dynamics 
and water maser dynamics (for a summary see Table 1s in Ho 1999 and Gebhardt \etal 2000a). 
Using the reverberation mapping technique, the SMBH masses of 20 Seyfert galaxies and 17 bright
quasars have been recently estimated (Ho 1999; Wandel \etal 1999; Kaspi \etal 2000). 
However, for most \sy galaxies, it is difficult to measure the SMBH mass using these 
methods because of either the large nuclear luminosity or the lack of long-term 
variability monitoring and precise measurements of characteristic velocity dispersions 
in the broad emission 
line region. 

%----------------------------------------------------------- Mbh-sigma
   \begin{figure}
   \centering
   \includegraphics[width=9.5cm, height=9.cm]{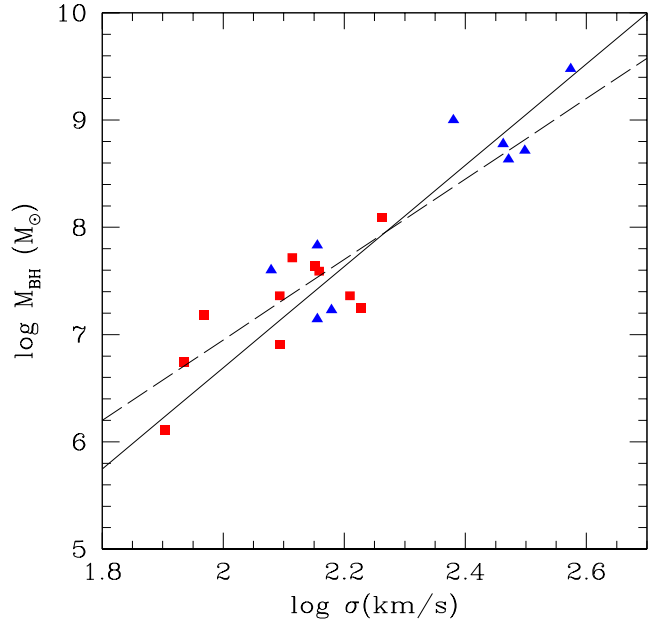}
      \caption{The black hole mass (in $\rm M_{\odot}$) %% 
against the velocity  dispersion for  %%
\sy galaxies. Squares and triangles represent the SMBH masses measured 
by reverberation mapping and stellar/gas dynamics, respectively. 
The solid and dashed lines correspond to the correlation found 
by Merritt \& Ferrarese (2000) and Gebhardt \etal (2000a), respectively.  
              }
         \label{Fig1}
   \end{figure}

Since the tight correlation between the 
SMBH mass and bulge velocity dispersion was found for
both normal and active galaxies (Gebhardt \etal 2000b; Ferrarese \etal 2001), it may be  
straightforward to estimate the SMBH mass 
from the measured bulge velocity dispersion. The correlation between the 
black hole mass and nuclear velocity dispersion for Seyfert galaxies
with both measured nuclear velocity dispersions and
black hole masses is demonstrated in Figure 1. 
The SMBH masses of 11 \sy galaxies (squares in Figure 1) 
were estimated by reverberation mapping (Wandel \etal 1999) and 
9 (triangles in Figure 1)
by the dynamical method (Ho 1999; Gebhardt \etal 2000a).
The values of their central velocity dispersions were taken from
Nelson \& Whittle (1995) and Ferrarese \etal (2001). 
Figure 1 clearly shows that \sy galaxies follow
the same $\rm M_{\rm BH}$-$\sigma$ relation as  normal galaxies (Gebhardt \etal 2000a; 
Merritt \& Ferrarese 2000). Although measurements of the
central velocity dispersions of these \sy galaxies were made by different groups, 
the small scattering around the
$\rm M_{\rm BH}$-$\sigma$ relation indicates that these measurements were reliable
and the systematic errors may not be important (Ferrarese \etal 2001). 
In this paper, we adopt the $\rm M_{\rm BH}$-$\sigma$ relation found by Merritt \& Ferrarese (2001), namely,
\be
\rm M_{\rm BH}=1.3 \times 10^8 \msun (\sigma/200~km~s^{-1})^{4.72}, %%
\ee
to derive the SMBH masses for 
\sy galaxies with measured nuclear velocity dispersions. 
Using a slightly flatter relation found by Gebhardt \etal (2000)  %%
does not cause significant changes in our results.

In the next sections we will compare some properties of \sy galaxies with those of
 quasars. Only a few quasars
 have SMBH masses determined by reverberation mapping. Laor (1998) %%
 adopted an empirical relation between  the size of the broad line region (BLR) and
the bolometric luminosity, $R_{BLR} \propto L^{1/2}$ (Kaspi \etal 1996; see also Kaspi \etal 
2000 for a slightly steeper relation), and derived the SMBH mass for
a number of quasars using the measured $H\beta$ velocity dispersion and the continuum luminosity.
We have compared the estimated SMBH masses for 9 \sy galaxies using
Laor's method (but adopted the BLR velocity $V=(\sqrt{3}/2)\rm{FWHM}(H\beta)$, as did  
Kaspi \etal (2000)) with the 
those obtained by reverberation 
mapping, and found that they agree well.
 This shows %% 
that using SMBH masses derived with different methods may not cause  serious 
problems in our present work.

%__________________________________________________________________
%__________________________________________________ two column table
 \tabcolsep 3.5mm
  \begin{table*}
      \caption[]{Black hole mass of a sample of Seyfert 1 and Seyfert 2 galaxies}
         \label{table1}
         \begin{tabular}{lcllrrrrcl}
            \hline
            %\noalign{\smallskip}         
Name &	Type &	$M_B^{tot}$ 	&	B-V	&  $M_B^{nuc}$ & $\rm{log} P^{nuc}_{6cm}$ &  $\rm{log} R_c$ &   $M_B^{bul}$ &  $\rm M_{\rm BH}$ &  	Note$^*$ \\
 & & & & & (W/Hz) & & & ($10^7 \rm M_\odot$)  & \\
  \hline
            %\noalign{\smallskip}
Mrk 279	&Sy1&	-20.92 &       0.69 &      -20.55 &      22.06&        0.16  &    -20.31 &       4.2$\pm$1.7	&	C,1,a\\
Mrk 335	&Sy1&	-21.48&        0.34&       -18.18 &      21.57&        0.62&      -20.62&        $0.63^{+0.23}_{-0.17}$	&	C,1\\
Mrk 590	&Sy1& -21.42&        0.67&       -16.46&       21.88&        1.62&       -20.40&        $1.78^{+0.44}_{-0.33}$ &	        C,1\\
Mrk 817	&Sy1&	-21.03&         0.40 &      -17.81&       22.02&        1.21&      -18.49 &       $4.4^{+1.3}_{-1.1}$	&	C,1\\
NGC 3227	&Sy1&  -20.57 &       0.82&       -16.01&        21.20&        1.12&      -19.55&      $3.9^{+2.1}_{-3.9}$	&	P,C,1\\
NGC 3516	&Sy1&	-20.63&        0.72  &     -17.21&       21.34&        0.78&      -20.02 &       $2.3\pm0.9$ &		P,C,1,a\\
NGC 4051	&Sy1&	-20.38&        0.67&       -14.97  &     20.54     &   0.87&      -18.41&        $0.13^{+0.13}_{-0.08}$	&	P,C,1\\
NGC 4151	&Sy1&	-20.16&        0.71  &     -19.18&       21.84&        0.49 &     -18.93&        $1.53^{+1.06}_{-0.89}$	&	P,C,1\\
NGC 5548	&Sy1&	-20.97&        0.62&       -17.29&       21.84 &       1.24&      -20.11&        $12.3^{+2.3}_{-1.8}$&		P,C,1\\
NGC 7469	&Sy1&	-21.32&        0.38&       -17.78&       22.43 &       1.64 &      -20.30&        $0.65^{+0.64}_{-0.65}$	&	C,1\\
NGC 3031	&Sy1&	-20.24 &       1.12   &    -11.73 &      20.16 &       1.79  &    -19.01   &      $0.68^{+0.07}_{-0.13}$	&	P,2\\
NGC 4258	&Sy1&	-20.13  &      0.77&      $>$-8.17 &      19.29 &      $>$2.34 &     -18.16&        $0.39\pm0.034$&		P,2\\
NGC 4395	&Sy1&	-16.51 &       0.53&      -8.69&       18.01&        0.85 &     $>$-16.51 &       $<0.008$	&	P,C,2\\
Mrk 530	&Sy1&	-21.48 &       0.72&      -16.27   &    22.24&        2.06&      -19.94&        $11.26\pm8.49$	&	C,3\\
Mrk 744	&Sy1&	-19.96 &       0.88 &     -17.56 &      20.94   &     0.24  &    -18.94&        $2.58\pm1.11$	&	C,3\\
NGC 1275	&Sy1&	-22.29  &      0.62   &   -18.53&       25.09     &      4.00 &     $>$-22.29&   $35.88\pm11.61$     	&	P,3\\
NGC 3982	&Sy1&   -19.43 &        0.80  &    -11.76&       20.16 &       1.78 &     -17.89&        $0.052\pm0.047$	&	P,3,b\\
NGC 4388	&Sy1&	-19.51 &        0.80 &     -13.17 &      21.19&        2.24  &    -17.97  &    $1.12\pm0.80$	&	P,3,d\\
NGC 4579	&Sy1&	-20.82   &     0.97&      -12.81&       21.16&        2.35 &     -19.28&       $6.04\pm3.02$	&	P,3\\
NGC 5252	&Sy1&	-20.93&           1.00  &    -14.23&        22.10 &       2.72 &     -20.32 &          $10.20\pm6.84$	&	C,3\\
NGC 5273	&Sy1&	-19.24 &       0.89  &    -13.51 &      19.98  &       0.90  &    -18.63 &       $0.16\pm0.08$	&	P,C,3\\
NGC 6104	&Sy1&	-21.12 &        0.80 &     -16.17&      $<$20.72 &       $<$0.57&       -20.10&        $3.14^{+3.50}_{-3.14}$	&	C,3,b\\
            %\noalign{\smallskip} 
 \hline

%==============
%Seyfert 2s: 15 objects
%
%Name     M_Btot       B-V       \Rm{Log} P6cm        M_B(bulge)     \Rm{Log} M_BH		Note
%
NGC 3079	&Sy2& -21.14  &        0.87  &  ...   &  21.97 & ...    &  -16.91    &     $1.3\pm0.4$ &			P,2\\
NGC 1068	&Sy2& 	-21.32 &        0.87  & ...    &  22.59 &  ...  &   -19.78  &       $1.7^{+1.3}_{-0.7}$	 &		P,C,2\\
NGC 1358	&Sy2& 	-20.95 &         1.05 & ...    &   21.56  &...   &   -20.09  &     $6.56\pm2.50$  	 &	P,3\\
NGC 1667	&Sy2& 	-21.52  &         0.80 & ...   &    21.99 & ...   &   -18.98  &     $6.56\pm4.65$   &			P,3,d\\
NGC 2273	&Sy2& 	-20.25  &        0.78  & ...  &    21.41  &...   &   -19.32   &       $1.36\pm0.52$	 &		P,3\\
NGC 3185	&Sy2& 	-18.99  &         0.80 &  ...   &   20.02  &...   &   -17.97 &        $0.048^{+0.074}_{-0.048}$		 &	P,3\\
NGC 5194	&Sy2& 	-20.76 &         0.91 & ...   &    20.17 & ...   &   -18.79 &        $0.54^{+0.63}_{-0.54}$	 &		P,3\\
NGC 7743	&Sy2& 	-19.78 &         0.91  & ...    &  20.33  &  ...  &  -19.05  &       $0.20^{+0.23}_{-0.20}$	 &		P,3\\
Mrk 573	&Sy2& 	-20.32   &       0.83   & ...    & 21.61   & ...  &  -19.71    &     $1.31\pm0.80$		 &	C,3\\
NGC 3362	&Sy2& 	-21.99  &         0.80   & ...  &    21.31  &... &     -19.45     &    $0.33^{+0.48}_{-0.32}$	 &		C,3,b\\
NGC 5929	&Sy2& 	-20.48  &         0.80  & ... &      21.46 &... &      -19.24    &     $1.21\pm0.62$ &			C,2,b\\
NGC 7682	&Sy2& 	-20.23   &        0.80  & ... &      22.08   &... &    -19.00      &   $1.31\pm0.85$	 &		C,3,b\\
NGC 5283	&Sy2& 	-19.40    &      0.92 & ... &       20.82 &... &      -18.79     &    $3.14\pm1.40$	 &		C,3\\
NGC 5695	&Sy2& 	-20.50    &      0.95  & ... &      19.91    &... &   -19.48  &       $2.76\pm0.99$	 &		C,3\\
NGC 7674	&Sy2& 	-21.77    &      0.83 & ... &      22.89   &... &    -19.81   &      $2.76^{+2.89}_{-2.76}$	 &		C,3\\

            %\noalign{\smallskip}
            \hline
         \end{tabular}
     
$^*$Notes: (C), object in CfA Seyfert sample; (P), object in Palomar \sy sample;
      (1), $\rm M_{\rm BH}$ measured by reverberation mapping; (2), $\rm M_{\rm BH}$ measured by gas/stellar dynamics;
      (3), $\rm M_{\rm BH}$ obtained in this paper using the $\rm M_{\rm BH}$--$\sigma$ relation; 
(a), uncertainty of   $\rm M_{\rm BH}$ not available in literature and assumed to be 40\%;
    (b), B-V value not available in literature and assumed to be 0.80; (d),  uncertainty of   $\sigma$ not available in literature and assumed to be 15\%. 
   \end{table*}

%__________________________________________________________________
\section{Sample of \sy galaxies}

We selected 37 \sy galaxies from two well-studied nearby \sy samples, 
the Palomar optical spectroscopic survey of bright ($B_T \leq 12.5 \rm mag$),
northern ($\delta > 0^o$) galaxies (Ho, Filippenko \& Sargent 1995),
including 21 \sy 1s and 28 \sy 2s (Ho \etal 1997a),
 the most complete and least biased available (Ho, Filippenko \& Sargent  1997b; %%
Ho \& Ulvestad 2001), and the CfA redshift Survey (Huchra \etal 1983) of 
galaxies with Zwicky magnitude $\leq 14.5$,
including 33 \sy 1s and 15 \sy 2s (Huchra \& Burg 1992; Osterbrock \& Martel 1993).
The \sy samples from these two surveys seem to complement one another,
though the combined sample may not be complete (Ho \& Peng 2001).

Our \sy sample include 22 \sy 1s and 15 \sy 2s in Palomar
and CfA samples  (see Table 1). These \sy galaxies have either 
the measured SMBH masses or the measured central velocity 
dispersions. Among them, three \sy 1s and two \sy 2s have dynamical
SMBH masses (Ho 1999;  Gebhardt \etal 2000a), and ten \sy 1s  
have SMBH masses measured by the reverberation mapping method 
(Wandel \etal 1999; Ho 1999). Another Nine \sy 1s and 13 \sy 2s have 
measured central velocity dispersions but unknown SMBH masses
 (Nelson \& Whittle 1995). The SMBH masses 
 can be estimated by equation (1) using the measured central velocity dispersions. 
Therefore, our sample consists of 
37 \sy galaxies with derived SMBH masses (see Table 1).

\section{Radio properties and black hole masses of Seyfert galaxies} %%
%

%                                                One column figure
%----------------------------------------------------------- R-Mbh
   \begin{figure}
   \centering
   \includegraphics[width=9.5cm, height=12cm]{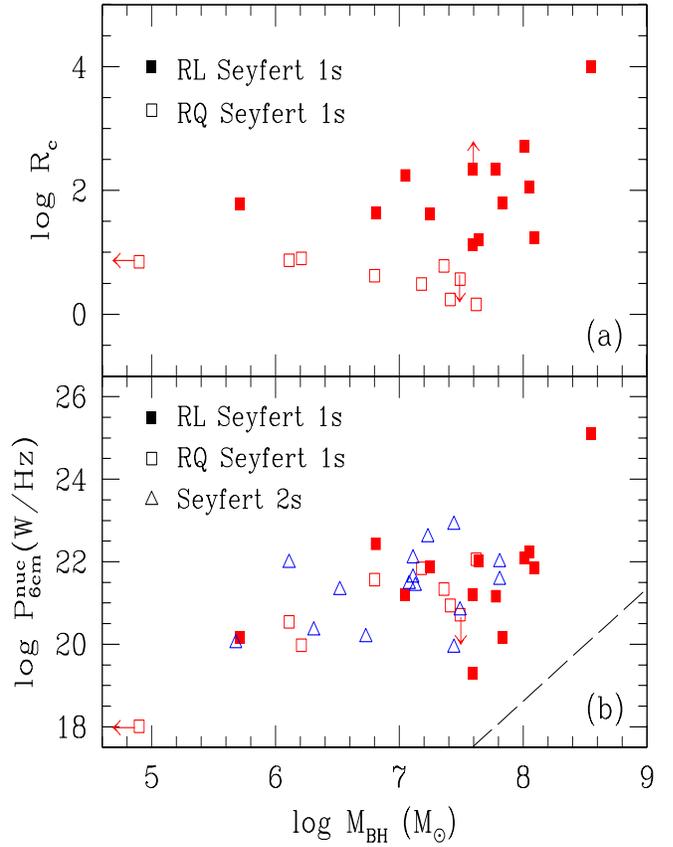}
      \caption{The nuclear radio-loudness and the nuclear radio power against the 
black hole masses for
		Seyfert galaxies. The dashed line in the lower panel represents the tight
		correlation between the core radio power and the SMBH masses  for nearby galaxies found by Franceschini \etal (1998).
              }
         \label{Fig2}
   \end{figure}
%
%______________________________________________________________

The radio properties of \sy galaxies have been investigated in detail using the Very Large Array (VLA)
at 3.6 cm by Kukula \etal (1995) for the CfA sample, and at 6 cm and 20 cm recently by Ho \& Ulvestad 
(2001) for the Palomar sample.  In this paper, we adopt the radio data for \sy 1s from Ho \& Peng (2001)
and the data for \sy 2s from Ho \&  Ulvestad (2001) and Kukula \etal (1995). The 6 cm data for
 some sources in the CfA sample (Kukula \etal 1995) were extrapolated from the 3.6cm data assuming 
$f_{\nu} \propto
\nu^{-0.5}$. 

\sy galaxies have been  considered usually as radio-quiet AGNs because most of them have lower 
radio-loudness, defined as the ratio of the radio to optical luminosities, 
 $R = L_{6 cm}/L_B$. However, a recently study of Ho \& Peng (2001) showed 
that though the nuclear radio power for \sy 1 galaxies on average accounts for
about $75\%$ of the total radio emission, the nuclear optical luminosity,
measured
by high-resolution optical images, accounts for merely 0.01\% of the integrated 
light. If the radio-loudness is measured by the nuclear
 radio and optical luminosities, 
most \sy 1s are in the category of radio-loud AGNs (with radio-loudness larger than 10). 
In Table 1, we listed the total and nuclear 
absolute B magnitudes ($M_B^{\rm tot}$ and $M_B^{\rm nuc}$)  and nuclear radio power ($P_{6cm}$) for %% 
\sy galaxies.   
The nuclear radio-loudness, $R_c$, is calculated by $R_c = L^{\rm nuc}_{6
cm}/L^{\rm nuc}_B$.  %%
The Hubble constant $H_0 =75 {\rm kms^{-1}Mpc^{-1}}$ and the deceleration parameter 
of $q_0=0.5$ were adopted. 
For \sy 2s, the measurements of their nuclear optical magnitudes are not available
 in the literature. 

Figure 2 shows the relation of SMBH masses with the nuclear radio-loudness
of \sy 1s and the 
radio power of \sy galaxies listed in Table 1. It is  clear  that the radio-loud \sy 1s 
with  larger nuclear radio-loudness seem to host more massive black holes. 
The similar tendency
has been found for PG quasars
by Laor (2000).  From Figure 2(b) we see 
a trend that \sy galaxies having a 
larger radio power perhaps host a more massive black hole than those of 
less radio power.
With the same black hole mass, \sy galaxies seem to have 100 to 1000 times
greater radio power than normal galaxies.   Laor (2000)
has shown that PG quasars also depart from such a correlation for nearby galaxies, with 
the radio luminosity of quasars being $10^4$ larger at a given $\rm M_{\rm BH}$. 
The difference of radio luminosities of  quasars, 
\sy galaxies and nearby galaxies may be simply due to different levels  %%
of nuclear activity. 

%__________________________________________________________________

\section{Black hole masses and bulge luminosities} %%

The correlation of black hole mass and galactic bulge luminosity found for nearby galaxies
implies a relationship 
between the black hole and bulge masses  
(Kormendy \& Richstone 1995; Magorrian \etal 1998; Ho 1999).  However, it is not clear
whether
the ratio between   black hole and bulge masses remains constant for all kinds of
 objects. Laor (2001) recently checked the correlation between  %%
the estimated black masses  and bulge luminosities for 
15 nearby quasars and 9 Seyfert galaxies, and found that they probably follow 
the same $\rm M_{\rm BH}$-bulge 
luminosity relation as for 16 nearby galaxies.
This suggests a universal nonlinear 
relation between the estimated black masses  and bulge luminosities,
${\rm M_{BH} \propto M}_{\rm bulge}^{1.54\pm0.15}$, for both normal and active 
galaxies. Therefore, ${\rm M_{BH} / M}_{\rm bulge}$ ratio is not constant for galaxies
with different bulge luminosities. However,  this 
needs to be confirmed by  larger
samples including both normal and active galaxies.

%                                                One column figure
%----------------------------------------------------------- Mvb-Mbh1
   \begin{figure}
   \centering
   \includegraphics[width=9.5cm, %height=12cm
				]{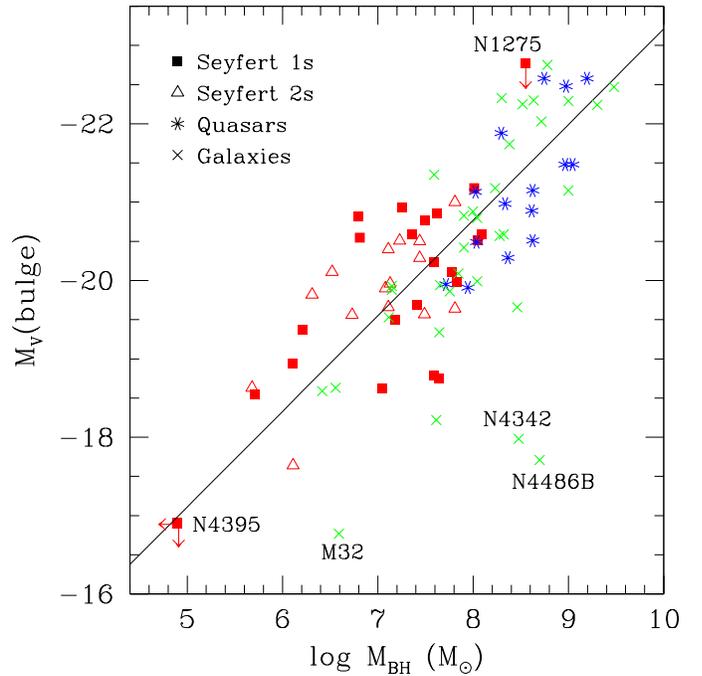}
      \caption{Correlation of V-band bulge absolute magnitudes and SMBH masses for Seyfert
galaxies, quasars and normal galaxies. The solid line represents the least $\chi^2$ fit 
to all objects by considering typical measurement errors of both parameters.
              }
         \label{Fig3}
   \end{figure}

\subsection{Our sample of \sy galaxies}

We use the 37 \sy galaxies with derived SMBH masses in our sample to
check the  ${\rm M_{BH} / M}_{\rm bulge}$ ratio. The absolute total B magnitudes  %%
(${\rm M}_{B}^{\rm tot}$) for our \sy sample (see Table 1) %%
were taken from Ho \etal (1997b) and Whittle (1992) (adapted to $H_0=75 {\rm kms^{-1}Mpc^{-1}}$).
The absolute bulge B magnitudes (${\rm M}_{B}^{\rm bulge}$) were obtained based on the 
relation between ${\rm M}_{B}^{\rm bulge}$ and ${\rm M}_B^{\rm tot}$ (Simien \& de Vaucouleurs 1986) and the Hubble stages %%
(defined in de Vaucouleurs, de Vaucouleurs \& Corwin 1976) of the host galaxies. The B-V colors were
taken from Ver\'on-Cettty \& Ver\'on (2000).
In order to compare our result with those obtained for quasars and nearby galaxies, we convert
$M_B^{\rm bulge}$ of \sy galaxies to the absolute bulge V magnitude, $M_V^{\rm bulge}$  (adapted to  %%
$H_0=80 {\rm kms^{-1}Mpc^{-1}}$)
by assuming that B-V color in the bulge is nearly the same as the total B-V color of \sy galaxy. 
In Figure 3 we show the relation between the SMBH masses and bulge V-band
 absolute magnitudes for \sy galaxies and 
the comparison of this relation
with those for quasars and nearby galaxies. The SMBH masses of 15 quasars and the absolute 
V magnitudes of their inner host were taken from Laor (1998, 2001).\footnote{Because
we used BLR velocity dispersion as $V=(\sqrt{3}/2)\rm{FWHM}(H\beta)$, the SMBH masses of
15 quasars are smaller  by a factor of 0.75 than those given by Laor (1998).
}
The SMBH masses determined by
stellar and gas dynamics for 33 nearby galaxies were taken from Kormendy \& Gebhardt (2001)
(we omitted M 81, a \sy 1 already in our \sy sample, and 3 objects with black hole masses  %%
measured by maser 
dynamics).
Their  absolute bulge
V magnitudes were derived from B-band absolute bulge magnitudes by adopting a standard
bulge color,  B-V = 0.94 (Worthey 1994).
It is evident that both \sy and nearby galaxies, as well as quasars, seem to follow the
same ${\rm M}_V^{\rm bulge}$-$\rm M_{\rm BH}$ relation. Note that unlike others, nearby galaxies  %%
NGC~4342, NGC~4486B and M~32 
deviate significantly
from the ${\rm M}_V^{\rm bulge}$-$\rm M_{\rm BH}$ relation. These offset galaxies are usually  fainter because %%
the outer regions of them may have been stripped away in the tidal interactions with
more massive companions (Faber 1973). These three galaxies, together with two \sy 1 galaxies, 
NGC 1275 with peculiar Hubble type and
NGC 4395 with only the upper limit of measured SMBH mass, are not included in our statistical studies. 
The measurement errors of $\rm M_{\rm BH}/M_\odot$ of our \sy sample were listed in Table 1, and 
those of normal galaxies were adopted from Kormendy \& Gebhardt (2001). For nearby quasars, such
errors are not available in Laor (1998) and assumed to be 60\%. The measurement errors of
${\rm M}_V^{\rm bulge}$ were seldom mentioned in the literature and thus are difficult to
 estimate. 
These errors for \sy galaxies were allocated to be 0.5, 0.75 and 1.0 mag respectively 
based on their quality assessment factors (Whittle 1992). Those of normal galaxies and nearby quasars
were adopted to be 0.5 and 0.75 mag, respectively.   
Taking into account these `typical' errors of ${\rm M}_V^{\rm bulge}$ and %% 
$\rm M_{\rm BH}$,
the least $\chi^2$ fit for 35 \sy galaxies, 15 quasars and 30 normal galaxies gives:
\be
M_V^{bulge} = -11.01 \pm 0.78 -(1.22 \pm 0.10 ) \rm{log} (\rm M_{\rm BH}/M_\odot).
\ee
The Spearman rank-order correlation coefficient is $r_S=-0.76$, which has a probability of 
$P_r = 3.3 \times 10^{-16}$ occuring by chance. Considering the errors in both parameters,
we adopted a bootstrap method to estimate the uncertainty of this correlation, and obtained
the mean correlation coefficient $\langle r_S \rangle=-0.63\pm0.05$.
Using the standard relation  %%
\be
M_V^{bulge} = 4.83 -2.5 {\rm{log}} {L_{bulge}/L_\odot}, 
\ee
and the relation between the bulge mass and luminosity (Magorrian \etal 1998), 
\be
 {\rm{log}} ({\rm M}_{\rm bulge}/{\rm M_\odot}) =-1.11+1.18 {\rm{log}} (L_{bulge}/L_\odot),
\ee
we then can get from eq. (2), 
$$
{\rm{log} ( M_{\rm BH}/M}_{\rm bulge}) =-11.06\pm1.11 +(0.74\pm0.14)~~~~~~~
$$
\be
~~~~~~~~~~~~~~~~~~~~~~~~~~~~{\rm{log} (M}_{\rm bulge}/
{\rm M_\odot}).
\ee
This gives 
${\rm M_{\rm BH} \propto M}_{\rm bulge}^{1.74 \pm 0.14}$. %%
The fitting for a sample of 35 \sy galaxies and 30 normal galaxies alone gives almost identical %%
results.
Laor (2001) recently found ${\rm M_{\rm BH} \propto M}_{\rm bulge}^{1.54 \pm 0.15}$ for  %%
a sample of objects including 
9 \sy galaxies and 15 quasars.
Our result shows that this nonlinear relation is more evident from a larger sample
including more \sy galaxies. 
Eq. (5) also shows that ${\rm M_{BH}/M}_{\rm bulge}$ is about 0.02\ %%%
when  ${\rm M_{BH}}=10^6{\rm M_\odot}$, and is 0.3\% when  ${\rm M_{BH}}=10^9{\rm M_\odot}$.
This clearly indicates that galaxies with more massive black holes have a larger
${\rm M_{BH}/M}_{\rm bulge}$ ratio. %%
%
%                                                One column figure
%----------------------------------------------------------- Mvb-Mbh2
   \begin{figure}
   \centering
   \includegraphics[width=9.5cm,
				]{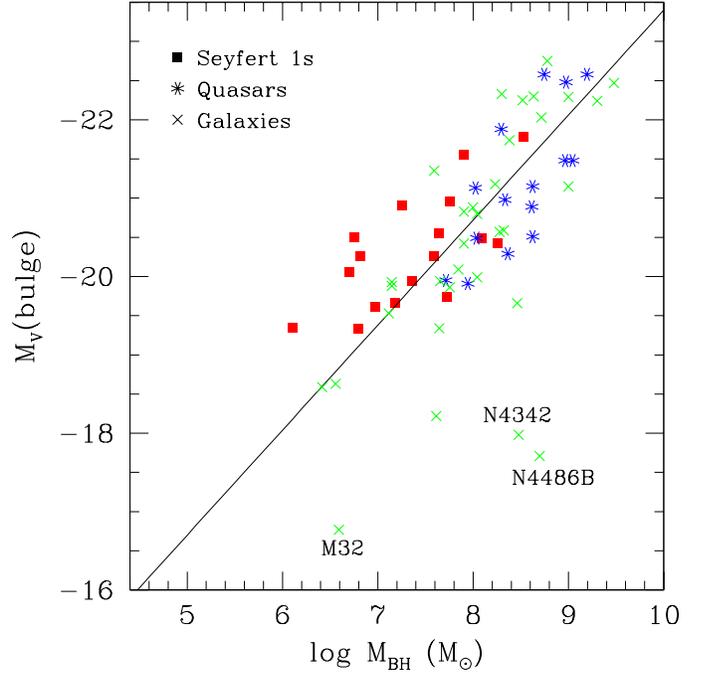}
      \caption{Correlation between the V-band absolute bulge magnitudes 
 and black hole masses for quasars,
 normal galaxies and Seyfert
1 galaxies with black hole masses only measured by reverberation mapping. 
The solid line 
represents the least $\chi^2$ fit 
to all objects by considering typical measurement errors of both parameters.
              }
         \label{Fig4}
   \end{figure}
%
%______________________________________________________________

\subsection{\sy sample in reverberation mapping studies}

We noted that the above 
 universal $\rm M_V^{\rm bulge}$-$\rm M_{\rm BH}$ relation (eq. 2) %%
is contrary to the result obtained by Wandel (1999), who found that
\sy 1 galaxies significantly depart from the  $\rm L_{\rm bulge}$-$\rm M_{\rm BH}$ relations of quasars and nearby galaxies %%
( Laor 1998; Magorrian \etal 1998).
However, Mclure \& Dunlop (2001) used the decomposed bulge luminosities of the same sample of  %%
 \sy galaxies 
as in Wandel (1999) and found no evidence 
for a different $\rm L_{\rm bulge}$-$\rm M_{\rm BH}$ %%
relation from quasars.
We noted that Wandel (1999) took the absolute bulge B magnitude, $\rm %%
 M_B^{\rm bulge}$, for \sy galaxies  %%
from Whittle (1992)(adopting $H_0=50 {\rm kms^{-1}Mpc^{-1}}$) and calculated the bulge luminosity
from $M_B^{bulge}$ with the  standard expression for $\rm M_V^{\rm bulge}$ (see eq. (3)) by assuming %%
$\rm M_B^{\rm bulge}\simeq M_V^{\rm bulge}$ for \sy 1s. %% 
We carefully checked these points and re-derived the absolute bulge V magnitude  of 17 \sy 1 galaxies with 
SMBH masses
measured by reverberation mapping (Wandel \etal 1999), using $H_0=80 {\rm kms^{-1}Mpc^{-1}}$ and assuming
the bulge  B-V color being nearly the same as the B-V color of the whole
 galaxy. The $\rm M_V^{\rm bulge}$-$\rm M_{\rm BH}$  %%
relation of 17 \sy 1 galaxies, as compared with those for 15 quasars (Laor 1998, 2001)
and 30
nearby 
galaxies with $\rm M_{\rm BH}$ measured by stellar dynamics (Gebhardt \etal 2000a), is shown in Figure 4.
Allocating the `typical' uncertainties of $\rm M_V^{\rm bulge}$  and  %%
$\rm M_{\rm BH}$ as we did above, the least $\chi^2$ fit for 62 objects gives,
\be
M_V^{bulge} = -10.00 \pm 0.96 -(1.34 \pm 0.12 ) \rm{log} (\rm M_{\rm BH}/M_\odot).
\ee
Considering the errors in both parameters,
we estimated the mean correlation coefficient as $\langle r_S \rangle=-0.63\pm0.06$.
This result is consistent with that
 found above
(see eq. (2)) and also indicates a nonlinear ${\rm M_{BH}}$-${\rm M}_{\rm bulge}$
relation.

%__________________________________________________________________
\subsection{Nelson \& Whittle's sample of \sy galaxies} 

In this section we explore the \sy sample in Nelson \& Whittle (1995), where the 
measurements of nuclear
velocity dispersions of about 70 \sy galaxies were reported. After excluding several LINERs
and normal galaxies, we got
33 \sy 1s and 32
\sy 2s. We estimated the SMBH masses of these \sy galaxies using the 
$\rm M_{\rm BH}$-$\sigma$ relation (eq. (1)). The {\it total} radio power at 5 GHz
of them was calculated from the 1.4 GHz data in Nelson \& Whittle (1995) by
assuming $f_\nu \propto \nu^{-0.5}$ and $H_0=80 {\rm kms^{-1}Mpc^{-1}}$. The radio-loudness
was calculated by the 5 GHz radio luminosity and the B-band optical luminosity.
The V-band bulge absolute magnitude was estimated from the B-band  bulge absolute 
magnitude (adapted to $H_0=80 {\rm kms^{-1}Mpc^{-1}}$) in Nelson \& Whittle (1995) by assuming
the B-V color of the bulge being the same as the total B-V color  (taken
from Ver\'on-Cetty \& Ver\'on (2000)) of \sy galaxy.

%                                                One column figure
%----------------------------------------------------------- Mvb-Mbh2
   \begin{figure}
   \centering
   \includegraphics[width=9.5cm, height=11cm
				]{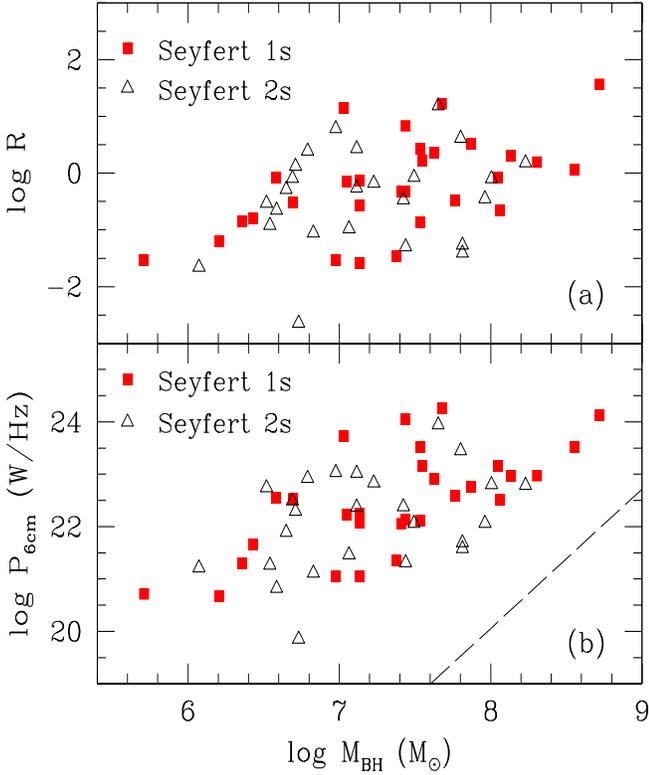}
      \caption{The radio-loudness (a) and radio power (b) against the black %%
   hole masses for \sy galaxies 
in the sample of Nelson \& Whittle (1995). The dashed line in (b) represents the tight
		correlation between total radio power and SMBH masses found by Franceschini \etal (1998) for nearby galaxies.
              }
         \label{Fig5}
   \end{figure}
%
%______________________________________________________________

%                                                One column figure
%----------------------------------------------------------- Mvb-Mbh2
   \begin{figure}
   \centering
   \includegraphics[width=9.5cm,
				]{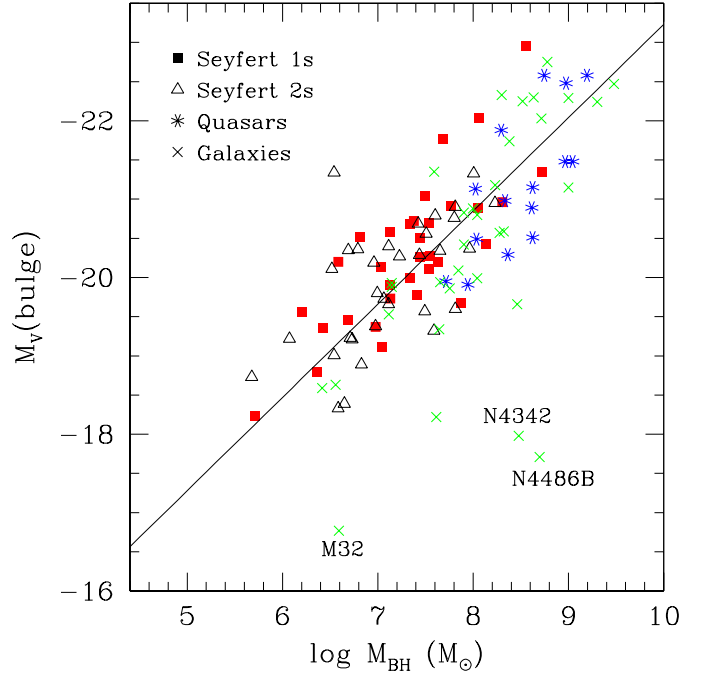}
      \caption{Correlation of V-band absolute bulge magnitudes and black hole masses for quasars,
 normal galaxies and Seyfert
galaxies in the sample of Nelson \& Whittle (1995). The solid line
represents the least $\chi^2$ fit
to all objects. The typical measurement errors of both parameters were considered. %%
              }
         \label{Fig6}
   \end{figure}

The radio-loudness and total radio power against the
SMBH masses for 29 \sy 1s and 25 \sy 2s with available radio data in the
sample of  Nelson \& Whittle (1995) are plotted in Fig.~5.
 Now we see 
that most \sy galaxies  seem to be radio quiet ($R<10$) when  we adopted the total 
radio and optical luminosities to calculate the radio-loudness. There is a weak
tendency that \sy galaxies with larger radio-loudness have larger SMBH masses. 
A comparison with the result in Figure 2(a) indicates that the nuclear radio-loudness
may be more fundamental and reflect the nature of central engine of \sy galaxies. 
From Figure 5(b) we see that there is a strong correlation between the total radio
power and the
SMBH mass, though the scatters are large. This confirms again the previous result that AGNs with larger radio power
may host more massive SMBHs (Franceschini \etal 1998; McLure \etal 1999). However, 
 as is shown in Figure 2,  \sy galaxies depart significantly from the tight relation
between the radio power and SMBH mass found for normal galaxies 
by Franceschini \etal (1998).

The relation between V-band bulge luminosities and SMBH masses for 33 \sy 1s and 32
\sy 2s in the sample of Nelson \& 
Whittle (1995) is shown in Figure 6. It again indicates a significant universal
relation between V-band bulge luminosities and SMBH masses for both \sy galaxies and 
quasars, as well as nearby galaxies. Adopting the `typical' uncertainties of 
${\rm M}_V^{\rm bulge}$ and  %%
$\rm M_{\rm BH}$ as we did above, the least $\chi^2$ fit for 65 \sy galaxies,
15 quasars and 30 nearby galaxies  gives,
\be
M_V^{bulge} = -11.33 \pm 0.66 -(1.19 \pm 0.09) \rm{log} (\rm M_{\rm BH}/M_\odot),
\ee
The simple Spearman rank-order correlation coefficient is $r_S=-0.76$, which has a probability of 
$P_r = 1.1 \times 10^{-21}$ occuring by chance.  Considering the errors in both parameters,
we estimated
the mean correlation coefficient as $\langle r_S \rangle=-0.61\pm0.04$.
This result is also almost identical to what
we found for the sample including 35 \sy galaxies. 
Therefore, the nonlinear $\rm{M_{BH}}$-$\rm{M_{\rm bulge}}$ relation
is confirmed by the 
substantially enlarged \sy sample.

\subsection{Effects of larger systematic errors}

A significant correlation between black hole mass and bulge luminosity was obtained
above by considering the `typical' errors of both parameters.
These errors are mainly due to the measurement uncertainties of bulge magnitude
and black hole mass. 
 However, the systematic errors are probably
substantially larger for both bulge magnitude
and black hole mass. For example, if $\rm M_{\rm BH}$ is determined by reverberation 
mapping, the systematic errors caused by the unknown BLR geometry, inclination may be as large
as a factor of 3 or more (Krolik 2001). Using the $\rm M_{\rm BH}$-$\sigma$ relation to estimate
$\rm M_{\rm BH}$ also has larger systematic errors due to the uncertainty of the slope of
this relation.
In addition, in deriving bulge magnitude from galaxy magnitude for \sy galaxies we adopted a 
statistical relation given by Simien \& de Vaucouleurs (1986). The systematic errors 
caused by applying this relation are
probably quite large and are difficult to  estimate quantitatively.

In order to estimate the effects of possible larger
systematic errors of black hole mass and bulge magnitude on our result, we adopted
a bootstrap method by adding the systematic errors into the uncertainties of
both parameters. 
Assuming the possible systematic errors for bulge absolute magnitude being 1 mag and %%
those for black hole mass being 90\% for all 35 \sy galaxies,  %%
15 quasars and 30 normal 
galaxies, we obtained the average correlation 
coefficient using the bootstrap approach as being $\langle r_S \rangle=-0.40\pm0.09$, which %%
indicates a moderately significant correlation. The minimum chi-square
fit considering both measurement and systematic errors in both parameter gives,
${\rm M}_V^{\rm bulge} = -10.49 \pm 1.77 -(1.29 \pm 0.22) \rm{log} (\rm M_{\rm BH}/M_\odot)$. %%
Comparing with the result we obtained in eq. (2), we found the 
slope of this relation  does not change very
much, though the uncertainty of the slope 
is doubled when the larger systematic 
errors are considered. The correlation coefficient  
also substantially decreases and its
uncertainty increases accordingly. However, these changes
are quite limited and have  no significant effects on the result we have obtained.
 %%Assuming the systematic errors of bulge magnitude larger than 1 mag will lead to
 %%a slightly shallower slope of the relation between bulge absolute magnitude
 %%and black hole mass, and hence a more significant non-linear relation
 %%between  black hole mass and bulge mass. Involving more large systematic
 %%errors in black hole masses will lead to a  steeper relation between bulge absolute magnitude
 %%and black hole mass but does not change our main results.

%______________________________________________________________
\section{Discussions}

Our results obtained for a larger sample of \sy galaxies show
 that AGNs %% 
with a larger nuclear radio-loudness
seem to have more 
massive black holes. This conclusion was obtained recently 
by Laor (2000)
for nearby quasars and was supported by our study for \sy galaxies.
Our results also strengthen the argument made by Ho \& Peng (2001) that the 
majority of \sy 1s are essentially radio-loud AGNs.
The total 
radio power of \sy galaxies increases with the black hole mass. 
At a given $\rm M_{\rm BH}$, quasars and \sy galaxies seem to
have greater radio power than that of nearby galaxies. These difference may  simply be due to the 
different level
of nuclear activity for different kinds of galaxies. 
A recent study using the quasars from the FIRST Bright Quasar Survey  found
evidence for the dependence of radio-luminosity on accretion rate and SMBH mass (Lacy \etal 2001). 
This may help us to understand the origin of scatters in the relation
 between the radio power and SMBH %%
mass.
If we describe the nuclear activity of galaxies
using the accretion rate $\dot{M}$, the difference of radio power at a given  $\rm M_{\rm BH}$ may show that quasars and \sy galaxies have larger $\dot{M}$  than nearby galaxies.
This seems also consistent with the picture that the accretion process in these systems may be 
different (Fabian \& Rees 1995; Di Matteo \& Fabian 1996).
Most likely
ADAFs with 
very low accretion rate exist in the nuclei of nearby 
galaxies, while quasars
and most AGNs probably host standard geometrically thin accretion disks with higher accretion rate.
However, the radio emissions from radio-quiet AGNs and nearby galaxies are  not well
understood at present. Whether they are from ADAFs (Narayan \etal 1998) or from weak jets 
(Falcke \& Biermann 1996, 1999) still remains uncertain. For radio-loud AGNs, the radio emissions
are thought to be mainly from the jet and are probably related to the magnetic fields or black
hole spin (Blandford \& Payne 1982; Blandford \& Znajek 1977). If the radio emissions correlate with
the black hole mass and accretion rate, we need to explain the possible relations of these
parameters with magnetic fields and black hole spin. However, no satisfactory theory can provide
 clear physics about these relations at present.     

Using the sample including 
 37 \sy galaxies in two well-defined bright \sy samples,
we studied 
the correlations of black hole masses
with V-band bulge absolute magnitudes and bulge masses
for active and normal galaxies. 
We find that the correlation 
between the black hole masses 
and the V-band absolute bulge magnitudes for Seyfert galaxies is almost consistent with 
those found for quasars and nearby galaxies. The combined sample of 37 Seyfert galaxies,
15 quasars and 30 nearby galaxies  seem to follow a universal nonlinear relation
between the black hole and bulge masses,
${\rm M_{BH} \propto M}_{\rm bulge}^{1.74 \pm 0.14}$, %%
which is slightly steeper
than that found recently by Laor (2001) for a sample including 
9 \sy galaxies. Our results
support the suggestion that the ratio of $\rm M_{\rm BH}$ and ${\rm M}_{\rm bulge}$ is not constant and galaxies  %%
with larger bulge luminosities or more massive SMBH probably have a
larger
${\rm M_{BH}/M}_{\rm bulge}$ ratio.  %%
Including  65 \sy galaxies in the larger sample of Nelson \& Whittle (1995)
led to almost the same result. 
In fact,
the nonlinear relation between $\rm M_{\rm BH}$ and ${\rm M}_{\rm bulge}$   %%
has been
 predicted  by some 
theories and  supported by some recent 
studies. For example, using a collapse model for black hole and bulge formation, Adams \etal (2001) predicted
${\rm M_{BH}/ M}_{\rm bulge}\propto \sigma$, which gives ${\rm M_{BH}/ 
 M}_{\rm bulge}\propto \rm M_{\rm BH}^{1/4}$ if  %%
$\rm M_{\rm BH} \propto \sigma^4$. This states clearly that the ratio of black hole and bulge masses
is not constant and can be larger for galaxies with more massive SMBHs. A similar result can be
obtained by exploring the model of Wang, Biermann \& Wandel (2000) who derived 
${\rm M_{BH}/ M}_{\rm bulge}\propto \sigma^{1.4}$. %%
In addition, recent studies on narrow line \sy 1s showed that their mean
 ${\rm M_{BH}/M}_{\rm bulge}$ %%
ratio is significantly smaller than that for normal \sy galaxies (Mathur, Kuraszkiewicz, \& Czerny 2001),
which  is also consistent with our results because narrow-line \sy 1s probably have smaller 
$\rm M_{\rm BH}$  than 
normal \sy galaxies (Boller, Brandt \& Fink 1996). 
 %%Other evidence for a nonlinear relation 
 %%between black hole and bulge masses
 %% has been given in Laor (2001).
 %%This nonlinear relation
 %%  has important implications on the theory
 %%of formation and evolution of black holes and galaxies. 

Finally we would like to mention that the tight correlation between the black hole mass and bulge luminosity %%
is not surprising and is actually expected from some existing well-known relations. 
From the $ {\rm M_{BH}}$-$\sigma$ relation and Faber-Jackson relation 
 between the bulge luminosity and $\sigma$
(Faber \& Jackson 1976), we certainly expect a relation between black hole mass and
bulge luminosity. The detailed studies of this relation and the 
$ {\rm M_{BH}}$-$ {\rm M}_{\rm bulge}$ relation are necessary because these relations are probably %% 
more fundamental than the Faber-Jackson relation and are more closely related to the 
physics of black hole and galaxy formation. 
We noticed that our result on the $ {\rm M_{BH}}$-$ {\rm M}_{\rm bulge}$ relation implies a rather  %%
flatter ${\rm L}_{\rm bulge}$-$\sigma$ relation than the Faber-Jackson %%
relation. This seems to be supported by the finding of  %%
 Nelson \& Whittle (1996).
 %%really found a shallow relation for \sy galaxies than that for  normal spiral bulges 
 %%and ellipticals.  
However, the slope of the ${\rm L}_{\rm bulge}$-$\sigma$ relation depends sensitively %% 
on the statistical method, sample selection 
and uncertainties of both parameters. In addition, in
deriving the $\rm M_{\rm BH}/M_{\rm bulge}$ ratio, we adopted the  %%
same bulge mass-to-light ratio for \sy
galaxies as for nearby galaxies obtained by Magorrian \etal (1998). 
Such a ratio may be smaller for \sy galaxies (Whittle 1992). Introducing a smaller bulge 
mass-to-light ratio may affect  our results.
We expect that further high quality observations on \sy galaxies,  normal
galaxies and quasars using the {\it Hubble Space Telescope} and larger ground-based telescopes 
could diminish the uncertainties in measuring the 
 galactic  bulge  properties and central black hole masses of these objects.
These efforts will undoubtedly help us to understand better the physics of formation and evolution
of SMBHs and galaxies.

\begin{acknowledgements}
      We are very grateful to the referee, Todd~Boroson, for helpful comments and  
suggestions. We thank Peter~Biermann, Xinwu~Cao, Jiansheng~Chen, Zugan~Deng, Jun~Ma, Tinggui~Wang,
Hong~Wu, Xiaoyang~Xia
 and Suijian~Xue for 
stimulating discussions. 
This work was partially supported by 
the Pandeng Project, the National Natural Science Foundation, and the
National Key Basic Research Science Foundation (NKBRSF G19990752)  in China.

\end{acknowledgements}

\end{document}